%% file: HFR.tex
\documentclass[twocolumn,prb,superscriptaddress]{revtex4-2}

\usepackage{amsmath,amsfonts,amssymb}
\usepackage{graphicx,color,xcolor}

\usepackage{hyperref}
\usepackage{epstopdf}
\usepackage{float}
\usepackage{multirow}
\usepackage{hhline}
\usepackage{ulem}
\usepackage{mathtools}
\usepackage{siunitx}
\usepackage[caption=false]{subfig}
\usepackage{soul}
\hypersetup{hidelinks,colorlinks=true,allcolors=BLACK}
\usepackage{blkarray}
\usepackage{amsmath}
\usepackage{textalpha}

\newcommand{\ket}[1]{\left|#1\right\rangle}

\DeclareSIPrefix\micro{\ensuremath{\mu}}{-6}

\makeatletter

\@ifundefined{textcolor}{}
{%
 \definecolor{BLACK}{gray}{0}
 \definecolor{WHITE}{gray}{1}
 \definecolor{RED}{rgb}{1,0,0}
 \definecolor{GREEN}{rgb}{0,1,0}
 \definecolor{BLUE}{rgb}{0,0,1}
 \definecolor{CYAN}{cmyk}{1,0,0,0}
 \definecolor{MAGENTA}{cmyk}{0,1,0,0}
 \definecolor{YELLOW}{cmyk}{0,0,1,0}
}

\usepackage[separate-uncertainty = true,multi-part-units=single]{siunitx}

\makeatother

\begin{document}

\title{Millimeter Wave Readout of a Superconducting Qubit}
% \title{Supplemental Material: Millimeter Wave Readout of a Superconducting Qubit}

\author{Akash V. Dixit}
\thanks{Present Address: Lawrence Berkeley Laboratory, Berkeley, CA 94720, USA}
\email{avdixit@berkeley.edu}
\affiliation{National Institute of Standards and Technology, Boulder, Colorado 80305, USA}

\author{Zachary L. Parrott}
\affiliation{National Institute of Standards and Technology, Boulder, Colorado 80305, USA}
\affiliation{Department of Physics, University of Colorado Boulder, Boulder, Colorado 80309, USA}

\author{Dennis Chunikhin}
\thanks{Present Address: University of Maryland, College Park, MD 20742, USA}
\affiliation{National Institute of Standards and Technology, Boulder, Colorado 80305, USA}

\author{Trevyn F. Q. Larson}
\thanks{Present Address: Google Quantum AI, Goleta, CA 93111, USA}
\affiliation{National Institute of Standards and Technology, Boulder, Colorado 80305, USA}
\affiliation{Department of Physics, University of Colorado Boulder, Boulder, Colorado 80309, USA}

\author{Bradley D. Hauer}
\affiliation{Institute for Quantum Computing and Department of Electrical and Computer Engineering, University of Waterloo, Waterloo, Ontario N2L 3G1, Canada}

\author{John D. Teufel}
\thanks{Present Address: AWS Center of Quantum Computing, Pasadena, CA 91106, USA}
\affiliation{National Institute of Standards and Technology, Boulder, Colorado 80305, USA}

% % \date{\today}
% % \date{March 10, 2026}

\begin{abstract}

Millimeter waves are emerging as an enabling technology for connecting and enhancing different quantum platforms such as Rydberg atoms, optomechanics, and superconducting qubits. In this work, we explore the interaction between millimeter wave photons and conventional transmon qubits, specifically for qubit readout. We study a circuit quantum electrodynamic (cQED) system consisting of a millimeter wave cavity at $\omega_r = 2\pi \times \SI{34.7}{\giga \hertz}$ and a transmon qubit at $\omega_q = 2\pi \times \SI{3.1}{\giga \hertz}$ coupled at rate $g = 2\pi \times \SI{1.3}{\giga \hertz}$. With such a large detuning between cavity and qubit, $\omega_r/\omega_q > 10$, we are able to suppress drive induced unwanted resonant state transitions, enabling strong drives for qubit readout. We measure no resonant state transitions even when driving with more than $100$ photons and read out the qubit state with more than $100$ photons to achieve a measurement fidelity greater than 99\% without the aid of a quantum limited amplifier.

\end{abstract}

\maketitle

\clearpage
\include{HFR_main_arxiv}
\include{HFR_supp}

\bibliography{HFR_references}

\end{document}

%% file: HFR_main_arxiv.tex
\section{Introduction}
There are many promising platforms for quantum information sciences, each with their strengths and weaknesses. By combining various technologies, we may be able to harness the best of each. Millimeter waves are emerging as an enabling technology to connect disparate quantum platforms such as circuit optomechanics, Rydberg atoms, and superconducting qubits. We envision a hybrid quantum system with capabilities such as state preparation and measurement from superconducting qubits, information storage from circuit optomechanics, and long range communication from optical transitions inherent to Rydberg atoms. Additionally, millimeter waves may unlock new frontiers for each of these individual quantum technologies \cite{kumar_quantum-enabled_2023, hauer_quantum_nodate, anferov_superconducting_2024, anferov_millimeter-wave_2025}.
% Extending the operating range of circuit QED devices into the millimeter wave range has already allowed for operation at higher temperatures \cite{anferov_superconducting_2024, anferov_millimeter-wave_2025}.

\begin{figure}
    \centering
    \includegraphics[scale=1.0]{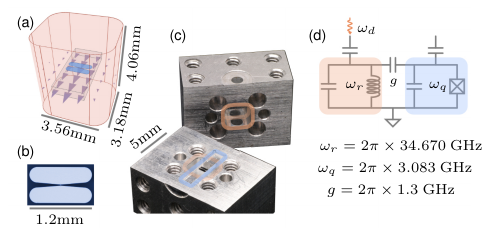}
    \caption{Circuit QED system with the transmon operating in the microwave frequency range and the readout resonator in the millimeter wave frequency range. (a) Transmon qubit coupled to electric field of millimeter wave frequency 3D cavity mode. The amplitude and direction of the simulated electric field are indicated by the size and orientation of the displayed arrows. (b) Optical micrograph of the transmon qubit used in this work. (c) The readout mode is implemented as the fundamental mode of a 3D cavity machined out of 6061 Al. The transmon is fabricated on a sapphire substrate with Al paddles and a Josephson junction (supplemental section \ref{supp:fab} \cite{noauthor_notitle_nodate}). (d) The readout mode and transmon are capacitively coupled at rate $g = 2\pi \times \SI{1.3}{\giga \hertz}$.}
    \centering
    \label{fig:device}
\end{figure}

In this work we aim to extend the toolbox of circuit QED into the millimeter wave range. This platform will allow us to study the interactions between conventional transmon qubits and highly detuned photons, which could lead to applications in quantum memories, error correction, and metrology at millimeter wave frequencies \cite{dixit_searching_2021, sivak_real-time_2023, li_cascaded_2025}. We explore how millimeter waves interact with conventional superconducting transmon qubits and demonstrate that millimeter waves provide protection against unwanted state transitions, a significant source of qubit readout error. We study a typical circuit QED system consisting of a transmon qubit and a readout resonator \cite{koch_charge-insensitive_2007}. Transmon qubits, usually in the frequency range of $\qtyrange[range-units=single, range-phrase={-}]{3}{6}{\giga\hertz}$, are typically read out using resonators in the $\qtyrange[range-units=single, range-phrase={-}]{6}{10}{\giga\hertz}$ range. However, when the transmon and drive frequencies are of the same order of magnitude, we can encounter various unwanted resonant transitions of the transmon state \cite{sank_measurement-induced_2016, shillito_dynamics_2022, khezri_measurement-induced_2023, cohen_reminiscence_2023, dumas_measurement-induced_2024, nesterov_measurement-induced_2024, wang_probing_2026, xia_exceeding_2025, dai_characterization_2026, li_mitigating_2025}. To mitigate these processes, we keep the transmon in the conventional frequency range while increasing the frequency of the readout resonator to the millimeter wave range (above $\SI{30}{\giga \hertz}$). We use the fundamental mode of a 3D cavity to host millimeter wave photons that interact with a transmon qubit, as shown in Figure \ref{fig:device}(a) \cite{paik_observation_2011}. With this system, we explore how millimeter waves may provide a route for improved qubit readout and establish millimeter wave photons as a resource for superconducting quantum systems.
% Additionally, millimeter waves may provide a route for improving qubit readout, a fundamental component of circuit QED systems.

% In the following sections, we present simulations showing the utility of performing qubit readout at millimeter wave frequencies. We describe the experimental protocol used to study potential state transitions in response to strong millimeter wave drives. This includes calibration of a multi-state transmon readout and the strong drives that we aim to study. We then present the probabilities of state transitions in response to drive photon numbers exceeding $1,000$ and find no induced resonant state transitions. Finally, we take advantage of the suppressed transition probabilities to perform qubit state readout with more than $100$ photons and achieve state measurement fidelity greater than 99\%.

\section{Numerical Simulations}
\begin{figure}
    \centering
    \includegraphics[scale=0.94]
    {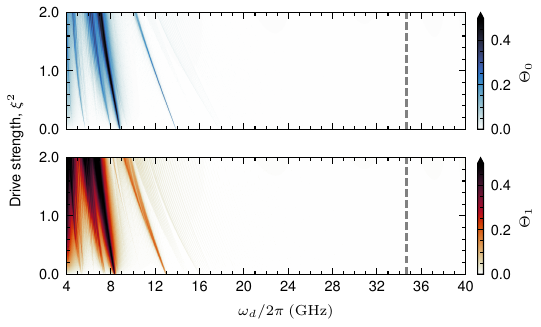}
    \caption{Floquet simulation of a charge driven transmon for varying drive frequency $\omega_d$ and normalized strength $\xi^2$ averaged over possible values of gate charge. The transmon's deviation from its ideal displaced state is shown for initial states $\ket{0}$ and $\ket{1}$ (upper and lower panels, respectively) over the range of typically chosen readout frequencies and extending into the millimeter wave range. The deviation is characterized by the parameter $\Theta_i$ indicating hybridization of the initial state with higher energy states of the transmon. We observe large hybridization and potential for state transition over much of the drive frequency and strength space of conventional qubit readout. With the readout drive at millimeter wave frequencies (above $\SI{30}{\giga \hertz}$), the simulations predict drive induced resonant state transitions are highly suppressed (See supplemental section \ref{supp:floquet} for more details \cite{noauthor_notitle_nodate}).}
    \label{fig:floquet_scars}
\end{figure}

% \cite{cohen_reminiscence_2023} does propose high freq readout as solutoin
% more photons means larger qubit state information rate or SNR
 Drive induced transitions of the transmon state can hinder the quantum nondemolition nature of the readout and limit the fidelity of state measurement. We explore readout with millimeter waves as a way to mitigate these state transitions. In Figure \ref{fig:floquet_scars} we a use Floquet analysis to simulate deviations from an ideal displaced state when a transmon, with frequency $\omega_q = 2 \pi \times \SI{3.083}{\giga \hertz}$ and anharmonicity $\alpha_q=-2\pi\times\SI{141}{\mega \hertz}$, is driven at various frequencies and strengths \cite{weiss_floquet_2024}. The deviations occur due to hybridization of the qubit states, $\ket{0}$ and $\ket{1}$, with higher transmon levels, increasing the probability of a state transition. This is characterized by the parameter $\Theta(\xi, \omega_d) = 1- \left| \left\langle j_{\text{Floquet}} | j_{\text{ideal}} \right\rangle \right|^2$, where  $\ket{j_{\text{Floquet}}}$ is the final state of the transmon as identified by the Floquet analysis and $\ket{j_{\text{ideal}}}$ is the ideal displaced state expected as a result of the drive. In the frequency range of a conventional readout, $\qtyrange[range-units=single, range-phrase={-}]{6}{10}{\giga\hertz}$, we find that even for low drive strengths there are many transitions, indicated by regions of larger hybridization, $\Theta$, making it difficult to avoid compromising the readout performance. These transitions correspond to resonances, at specific drive frequencies and strengths, that transfer population from the computational states of the qubit to higher energy transmon levels. Since the higher energy transmon states are highly susceptible to charge noise, the transitions can be activated over a broad range of drive frequencies and strengths, making it difficult to engineer a window free of transitions \cite{dai_characterization_2026}. Recent work has shown that engineering alternative coupling structures or implementing a large detuning between the readout mode and transmon can eliminate such state transitions \cite{chapple_balanced_2025, kurilovich_high-frequency_2025, connolly_full_2025, mori_suppression_2026, beaulieu_fast_2026}.

The strategy of large detuning is motivated by the behavior of the transmon charge matrix elements, which determine the probability of drive induced state transitions. The charge matrix elements connecting the ground or first excited state to higher levels (of opposite excitation number parity) are relatively large for final states bound within the cosine potential of the Josephson junction, but are exponentially suppressed when the final state lives outside the confines of the potential well. To avoid connecting states with high probability, we should avoid drives that match the frequencies of transitions between states within the potential well. For typical circuit QED experiments this can be done by operating with drive frequencies at least 10 times the ground to first excited state transition (see supplemental Figure \ref{fig:charge_states} \cite{noauthor_notitle_nodate}).

% Resonant transitions between initial and final states can be activated with a relatively weak drive, at the difference frequency of the initial and final states, if the corresponding charge matrix element is large.
% % To avoid transitions, we should not drive at frequencies that connect initial states to final states inside the Josephson junction potential where the relevant charge matrix elements are relatively large.
% Though it is difficult to completely avoid any drive frequency that connects transmon states, we can ensure that a drive only connects states where the charge matrix elements are heavily suppressed by choosing drive frequencies greater than the height of the Josephson junction potential well. For typical circuit QED device parameters, there are about 10 states bound in the potential well, implying that we should operate with drive frequencies at least 10 times larger than the frequency of the ground to first excited state transition.

The most common example of such a drive is the readout tone used to probe the state of the transmon. We can mitigate readout induced unwanted resonant state transitions by operating with a large detuning between the transmon and readout frequencies, $\omega_r/\omega_q>10$. This was recently achieved by lowering the transmon frequency \cite{kurilovich_high-frequency_2025}. In this work, we achieve a large readout-transmon frequency ratio, $\omega_r/\omega_q>10$, by increasing the readout frequency to the millimeter wave range and keeping the transmon in the microwave frequency range, as shown in Figure \ref{fig:device}(b) \cite{mencia_raising_2025}. This allows us to maintain the required transmon anharmonicity needed for fast gates and operate with minimal residual thermal population. An additional benefit could be reduced thermal occupancy of the readout mode for typical cryostat temperatures, leading to reduced qubit dephasing. The simulations presented in Figure \ref{fig:floquet_scars} show that for readout drives in the millimeter wave frequency range $(> \SI{30}{\giga \hertz})$ resonant state transitions are highly suppressed.
%maybe add something about typical 3-5GHz qubit mataining large anharmonicty for fast gates and lower thermal poulation for a given cryostat temperature
% higher freq readout could result in lower residual population and longer qubit coherence

\section{Experimental Setup and Protocol}
We operate a superconducting transmon qubit whose transition frequency is in the microwave frequency range $\omega_q= 2\pi\times\SI{3.083 }{\giga \hertz}$ with an anharmonicity of $\alpha_q = -2\pi \times \SI{141}{\mega \hertz}$. We read out the state of the transmon using the fundamental mode of a 3D Al cavity at millimeter wave frequencies, $\omega_r= 2\pi\times\SI{34.670}{\giga \hertz}$. The device is mounted to the base plate of a dilution refrigerator at $\SI{10}{\milli \kelvin}$. The transmon and readout mode shown in Figure \ref{fig:device} are dispersively coupled with the interaction Hamiltonian $\mathcal{H}_I \supset \chi a^{\dagger}_q a_q a^{\dagger}_r a_r$. $a_q, a^{\dagger}_q$  and $a_r, a^{\dagger}_r$ are the annihilation and creation operators for the transmon and readout mode, respectively. $\chi$ is the cross Kerr interaction rate between the transmon and the readout mode. We measure $\chi = -2\pi\times\SI{1.515}{\mega \hertz}$, achieved by a readout-transmon coupling rate of $g = 2 \pi \times \SI{1.3}{\giga \hertz}$ (see Figure \ref{fig:device}(b)). When the transmon is in state $\ket{n}$, the readout resonance frequency shifts by approximately $n\chi$ (see supplemental section \ref{supp:device_params} \cite{noauthor_notitle_nodate}). The readout linewidth is $\kappa = 2\pi \times \SI{1.997}{\mega \hertz}$, the transmon energy relaxation time is $T_1= \SI{110}{\micro \second}$, and echo coherence time is $T_2^E = \SI{188}{\micro \second}$.

We test the resilience to resonant state transitions due to highly detuned readout by probing the transmon response to strong drives. We first prepare and read out the transmon state and then apply a drive near the resonant frequency of the readout mode. Since resonant state transitions occur at a specific combination of drive frequency and drive strength, we search for potential transitions by varying the number of drive photons applied. Finally, we read out the transmon state. A change in the state from the first to the second measurement is the sign of a state transition. A change in the transition probability in response to a particular drive strength at the chosen drive frequency would indicate a resonant state transition.

To determine which state the transmon has potentially transitioned to, we set up our readout to differentiate multiple states. We probe at a frequency between the resonator frequency when the transmon is in the first or second excited state ($\omega_r + 2\chi < \omega < \omega_r + 1\chi$, see supplemental Figure \ref{fig:cross_kerr} \cite{noauthor_notitle_nodate}). At this frequency, we read out with $\bar{n}_r = 8$ for $\tau_r = \SI{10}{\micro \second}$ to distinguish between transmon states $\ket{0}, \ket{1}, \ket{2}, \ket{3}$. The same multistate readout is used to determine the transmon state before and after applying the drive we are testing. See supplemental Figure \ref{fig:multistate_readout} for more information about the multistate readout \cite{noauthor_notitle_nodate}.

We test drives that can potentially induce state transitions by pumping slightly detuned from the readout frequency, $\omega_d = \omega_r + \Delta_d$, with a $\tau_d = \SI{1}{\micro \second}$ pulse. We calibrate this drive by measuring the first transmon transition frequency, which will shift in response to the drive, an effect known as the AC-Stark shift \cite{schuster_ac_2005}. The shift is proportional to the effective drive photon number, $\Delta \omega_q = \bar{n}_d \chi$. We measure the AC-stark shift when the transmon is in state $\ket{0}$ to directly calibrate drive photon numbers up to around $\bar{n}_d = 175$. We then fit the relationship between normalized drive amplitude ($A$) and drive photon number with the expression $\bar{n}_d(A) = C A^2 \kappa / \left[(\Delta_d -K \bar{n}_d(A) - n \chi)^2 + (\kappa/2)^2 \right]$ to obtain the scaling factor, $C$, and the self-Kerr shift of the readout mode, $K$. This implicit function accounts for the drive detuning due to the off resonant drive frequency ($\Delta_d = 2\pi \times \SI{1.626}{\mega \hertz}$), the transmon state dependent shift $(n \chi)$, and the inherited nonlinearity of the readout mode, characterized by its self-Kerr shift $(K\bar{n}_d(A))$. We extrapolate this relationship to calibrate the drive to photon numbers approaching $\bar{n}_d =1000$ as a function of the drive amplitude for all initial transmon states (see supplemental Figure \ref{fig:stark_shift} \cite{noauthor_notitle_nodate}).

% The shift is linear in the effective drive photon number $\Delta \omega_q = \bar{n}_d \chi$. We calibrate the relationship between drive strength and drive photon number up to $\bar{n}_d = \alpha_q/\chi \sim 100$ using the AC-Stark shift, see supplemental Figure \ref{fig:stark_shift}. We extrapolate to larger $\bar{n}_d$ using the quadratic relationship between drive strength and drive photon number calibrated at low drive strengths. In order to explore large induced photon numbers we choose a small drive detuning of $\Delta_d = 2\pi \times \SI{1.626}{\mega \hertz}$. We keep the drive frequency $\omega_d$ fixed for all prepared initial states of the transmon. Since we use a positive detuning and the cross Kerr shifts are negative, the resonator response shifts away from the drive by $n\chi$ when the transmon is prepared in $\ket{n}$. For a given drive strength applied to the system, this effectively means that the number of induced photons in the resonator is smaller for larger initial states $\ket{n}$. We correct for this effect by scaling the number of drive photons by a factor dependent on the initial state. The correction takes the form of a ratio of Lorentzian responses $\bar{n}_d(\ket{n}) = \bar{n}_d(\ket{0}) \frac{\Delta_d^2 + \kappa^2}{(\Delta_d - n\chi)^2 + \kappa^2}$.

\section{Results and Discussion}
% \begin{figure}
%     \centering
%     \includegraphics[scale=1.0]{Figures/transition_prob_01.pdf}
%     \caption{Experimental data probing drive induced transitions of the transmon with $\omega_d/\omega_q=11.25$. We probe the effect of a drive near the readout frequency with varying strength on the transmon qubit state. The transmon is prepared in either (a) $\ket{0}$ or (b) $\ket{1}$ 20,000 times. We track the state of the transmon after applying a variable strength drive with photon number $\bar{n}_d$ and measure the probability of transitioning from the initial state to various final states. With the transmon initialized in the $\left\{ \ket{0}, \ket{1} \right\}$ subspace, we find no observable resonant state transitions. At drive photon numbers $\bar{n}_d > 100$, we observe the onset of the quantum to classical regime. The transmon state is distributed over many levels as seen by the rise in the probability of transition to the $\{\ket{n}\}_{n\geq4}$ manifold while initialized in either $\ket{0}$ or $\ket{1}$.}
%     \centering
%     \label{fig:transition_prob_01}
% \end{figure}

\begin{figure}
    \centering
    \hspace{-0.8cm}
    \includegraphics[scale=1.0]{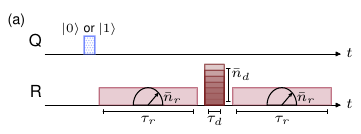}
    
    \vspace{-0.25cm}
    
    \includegraphics[scale=1.0]{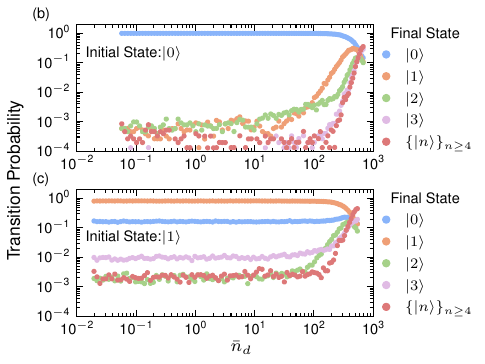}
    \caption{Experimental data probing drive induced transitions of the transmon with $\omega_d/\omega_q=11.25$. (a) The transmon is prepared in either $\ket{0}$ or $\ket{1}$, we apply a drive near the readout frequency with varying strength, and record the transmon state before and after the drive. The transmon is prepared in either (b) $\ket{0}$ or (c) $\ket{1}$ 20,000 times each. We track the state of the transmon after applying a variable strength drive with photon number $\bar{n}_d$ and measure the probability of transitioning from the initial state to various final states. With the transmon initialized in the $\left\{ \ket{0}, \ket{1} \right\}$ subspace, we find no observable resonant state transitions for drive photon numbers up to $\bar{n}_d = 100$. At drive photon numbers $\bar{n}_d > 100$, we observe the transmon escape the confined of its potential well. The transmon state is distributed over many levels as seen by the rise in the probability of transition to the $\{\ket{n}\}_{n\geq4}$ manifold while initialized in either $\ket{0}$ or $\ket{1}$.}
    \centering
    \label{fig:transition_prob_01}
\end{figure}

In Figure \ref{fig:transition_prob_01} we show the probabilities of transmon qubit state transitions from an initial state $\ket{0}$ or $\ket{1}$ to all the measurable final states $\ket{0}, \ket{1}, \ket{2}, \ket{3}, \{\ket{n}\}_{n\geq4}$ as a function of the number of drive photons. $\{\ket{n}\}_{n\geq4}$ is the subspace of transmon states with four or more excitations. A state transition would be identified as a reduction in probability of the transmon to remain in its initial state and a corresponding increase in the transition probability to a different final state. For resonant state transitions, this typically presents as a peak around or onset at a particular drive photon number \cite{sank_measurement-induced_2016, wang_probing_2026, dai_characterization_2026}.  We find that there are no measurable drive induced resonant transitions to states outside the qubit subspace while starting initially in either $\ket{0}$ or $\ket{1}$ for drive photon numbers up to $\bar{n}_d = 100$. To quantify this, we use a one-tailed test for deviations from the average probability to stay in the initial state for drive photon numbers up to $\bar{n}_d = 100$, $\bar{P}_{\ket{n}\rightarrow\ket{n}}$. When the transmon is initialized in state $\ket{0}$, $\bar{P}_{\ket{0}\rightarrow\ket{0}} = 0.998$, consistent with a residual qubit population of $\bar{n}_q=0.012$, and when initialized in state $\ket{1}$, $\bar{P}_{\ket{1}\rightarrow\ket{1}} = 0.821$, consistent with a total experiment time of $2 \tau_r + \tau_d = \SI{21}{\micro \second}$ and qubit $T_1 = \SI{110}{\micro \second}$. We rule out deviations, in the probability of leaving the initial state, greater than 0.0014 and 0.0012 with a confidence level of 95\% for initial states $\ket{0}$ and $\ket{1}$, respectively. The lack of any resonant features connecting the initial state to any other transmon state in both Figure \ref{fig:transition_prob_01}(a) and (b) are consistent with the simulations in Figure \ref{fig:floquet_scars}, showing that resonant transitions are highly suppressed for the large transmon-readout detuning we have engineered, $\omega_r/\omega_q > 10$. Supplemental section \ref{supp:cavity_mediated_transition} shows the transition probabilities when the transmon is initialized in higher levels $\ket{2}$ or $\ket{3}$, where we also find no resonant intra-transmon transitions for drive photon numbers up to $\bar{n}_d = 100$, but observe transitions mediated by a higher order mode of the 3D cavity \cite{noauthor_notitle_nodate}.

After a large drive, with drive photon numbers $\bar{n}_d > 100$, the transmon escapes the confinement of its potential well, and we find the population distributed over many states. This is observed as an increase in the transition probability to the $\{\ket{n} \}_{n \geq 4}$ manifold, consistent with numerical simulations in supplemental Figure \ref{fig:QtoC} \cite{pietikainen_observation_2017, bishop_response_2010, boissonneault_improved_2010, lescanne_escape_2019, noauthor_notitle_nodate}. This phenomenon is inherent to dispersively coupled circuit QED systems and is the ultimate limit on the applied drive strength.

In the absence of observable resonant state transitions, we are able to resolve weaker forms of drive induced leakage in Figure \ref{fig:transition_prob_01}. For example, we identify a form of inelastic scattering where the drive adds two transmon excitations and the remaining energy is emitted as an off resonant photon into the environment \cite{connolly_full_2025}. In Figure \ref{fig:transition_prob_01}(b) the relevant process is shown in the green curve, where $P(\ket{0} \rightarrow\ket{2})$ scales as $\bar{n}_d$. We also see potential signatures of quasiparticle induced transitions where multiple drive photons break a Cooper pair and either remove or add an excitation to the transmon \cite{chowdhury_theory_2026, kishmar_quasiparticle-induced_2026}. In Figure \ref{fig:transition_prob_01}(b) the relevant process is shown in the orange curve where $P(\ket{0} \rightarrow\ket{1})$ scales as $\bar{n}_d^3$. The processes that have been identified are the subject of further study and other forms of state leakage should be explored in the context of highly detuned millimeter waves impinging on a transmon. Future work could pursue strategies such as carefully engineering the dissipative environment seen by qubits to reduce inelastic scattering and choosing a slightly lower drive frequency to reduce the generation of quasiparticles while maintaining the protection against resonant state transitions.

\begin{figure}[t]
    \centering
    \includegraphics[scale=1.0]{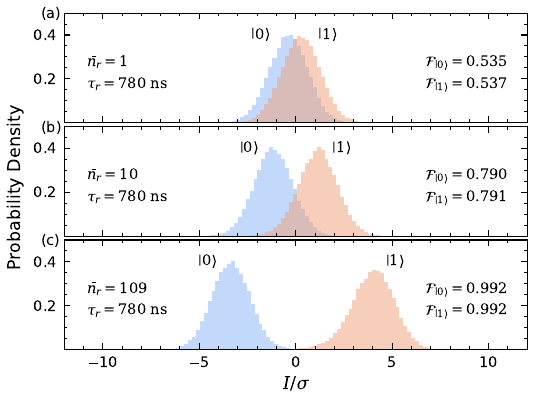}
    \caption{Experimental data of reading out qubit state with increasing photon number. We prepare the qubit state in $\ket{0}$ and $\ket{1}$ 50,000 times and readout for a fixed time of $\tau_r = \SI{780}{\nano \second}$. As we increase the number of readout photons we observe an improvement in the measurement fidelity, (a) $\bar{n}_r = 1, \mathcal{F}_{\ket{0}} =\mathcal{F}_{\ket{1}} = 0.535$, (b) $\bar{n}_r = 10, \mathcal{F}_{\ket{0}} =\mathcal{F}_{\ket{1}} = 0.790$, (c) $\bar{n}_r = 109, \mathcal{F}_{\ket{0}} =\mathcal{F}_{\ket{1}} = 0.992$. After the $\tau_r = \SI{780}{\nano \second}$ readout is complete, it takes time $T = \ln{\left(\bar{n}_r/\bar{n}_0^r\right)/\kappa}$ for the resonator to relax to its residual population $\bar{n}_0^r = 1.06 \times 10^{-3}$. For $\bar{n}_r \in \left\{1,10,109 \right\}$, $T \in \left\{\si{546}, \si{729}, \si{920} \right\} \si{\nano \second}$.}
    \centering
    \label{fig:fidelity}
\end{figure}

We harness the ability to drive the readout resonator with large photon number, without inducing resonant transitions of the transmon state, to perform high power qubit readout. In Figure \ref{fig:fidelity} we show a projection of the distribution of recorded in phase and quadrature signals ($I,Q$) for the qubit prepared in $\ket{0}$ or $\ket{1}$ while varying the number of photons that populate the readout resonator. We calibrate the resonator photon number using the AC-Stark effect and readout with $\bar{n}_r \in \left\{1,10,109 \right\}$ for a fixed time of $\tau_r = \SI{780}{\nano \second}$. In $I,Q$ space we use a circular separatrix centered at the median of the $\ket{0}$ distribution to label each measurement either $0$ or $1$. The state measurement fidelity is given by $\mathcal{F}_{\ket{n}} = P(n|\ket{n})$, the probability of labeling a measurement with 0 or 1 given the initial state is correspondingly $\ket{0}$ or $\ket{1}$. We optimize the size of the circle to simultaneously maximize $\mathcal{F}_{\ket{0}}$ and $\mathcal{F}_{\ket{1}}$, see supplemental Figure \ref{fig:state_assignment} for details \cite{noauthor_notitle_nodate}. With $\bar{n}_r = 109$ we achieve a measurement fidelity of $\mathcal{F}_{\ket{0}} = \mathcal{F}_{\ket{1}} = \num{0.992}$.

Further improvements to measurement fidelity could be achieved by improving the qubit decay time and reducing the readout time to reduce decay and heating induced infidelity. Optimization of state preparation and state reset would increase the probability of starting in the intended state. The measurement efficiency of the readout chain is $1.74 \%$ corresponding to a system added noise of $\bar{n}_{\mathrm{sys}} = 28$ photons, see supplemental Figure \ref{fig:efficiency} for details \cite{gambetta_quantum_2008, clerk_introduction_2010, noauthor_notitle_nodate} . Developing and adding a quantum limited amplifier at millimeter wave frequencies would greatly improve the measurement efficiency \cite{macklin_nearquantum-limited_2015, hao_wireless_2026}.
% \TL{Is it worth estimating a readout fidelity at at say nsys=2? }

\section{Conclusion}
In this work, we demonstrate the utility of expanding the frequency range of circuit QED experiments into the millimeter wave regime. We demonstrate the feasibility of operating cQED experiments with qubits and photons of disparate frequencies. In this regime, we achieve a strong dispersive interaction between qubit and resonator with a coupling rate an order of magnitude larger than for typical cQED experiments, while maintaining the coherence of the qubit. We show that resonant state transitions of the transmon state can be mitigated by highly detuning the readout mode from the transmon, $\omega_r/\omega_q>10$. Because the drive induced state transitions are suppressed, we are able to perform a high fidelity qubit readout using more than $100$ photons.

Looking forward, millimeter waves may have applicability for superconducting qubit based processors and their feasibility for fast, multiplexed readout in 2D systems should be extensively studied. In addition to suppressing intra-qubit resonances, a large frequency separation between qubits and their readout resonators could minimize accidental collisions with parametric processes used for inter-qubit connections.

% As superconducting qubit based processors scale to larger numbers of qubits, the number of drives required for control and readout will also increase. In addition to suppressing intra-qubit resonances, a large frequency separation between qubits and their readout resonators could minimize accidental collisions with parametric processes used for inter-qubit connections.

This initial work coupling millimeter waves and transmon qubits paves the way for further advancements in circuit QED experiments. This includes studying 3D bosonic systems at millimeter wave frequencies, to enable quantum memories and error correction \cite{kuhr_ultrahigh_2007, chakram_multimode_2022, sivak_real-time_2023, li_cascaded_2025}. This platform extends the techniques of quantum metrology to the millimeter wave range, useful, for example, in dark matter searches \cite{dixit_searching_2021, agrawal_stimulated_2024}. The coupling between millimeter wave photons and qubit demonstrated here may allow us to further explore ultrastrong light matter coupling in circuit QED systems \cite{frisk_kockum_ultrastrong_2019}. More broadly, this work is another example of the utility that millimeter waves bring to various quantum information platforms \cite{kumar_quantum-enabled_2023, hauer_quantum_nodate}. Millimeter wave photons may be a central technology for hybrid quantum systems that leverage the capabilities of platforms such as circuit optomechanics, Rydberg atoms, and superconducting qubits.

\section*{Acknowledgments}
The authors acknowledge and thank Martin Gould for machining the 3D cavity, Daniel Weiss, Martin Ritter, Shawn Geller, and Aziza Suleymanzade for useful discussions, and Adam Sirois for providing support for initial millimeter wave testing.

The numerical simulations in this work were made possible by the large memory analysis system \texttt{postproc-b} on NIST's Boulder campus and supported by NIST's Research Services Office.

Certain commercial instruments are identified to specify the experimental study adequately. This does not imply endorsement by NIST or that the instruments are the best available for the purpose.

\clearpage

%% file: HFR_supp.tex
\clearpage
\newpage
\renewcommand{\thesection}{S\Roman{section}}
\renewcommand{\thefigure}{S\arabic{figure}}
\renewcommand{\thetable}{S\arabic{table}}
\renewcommand{\theequation}{S\arabic{equation}}
\renewcommand\thepage{S\arabic{page}}
\setcounter{figure}{0}
\setcounter{table}{0}
\setcounter{equation}{0}
\setcounter{section}{0}
\setcounter{page}{1}
% \global\long\def\theequation{S\arabic{equation}}
% \global\long\def\thefigure{S\arabic{figure}}
% \global\long\def\thetable{S\arabic{table}}

\DeclareSIUnit{\torr}{Torr}
    
\onecolumngrid
\begin{center}
    % \vspace{5ex}
    \centerline{\large \textbf{Supplemental Material:}}
    \vspace{2ex}
    \centerline{\large \textbf{Millimeter Wave Readout of a Superconducting Qubit}}
    % \vspace{5ex}
    % \normalsize Akash V. Dixit, Zachary L. Parrott, Dennis Chunikhin, \\
    % \vspace{0.5ex}
    % Trevyn F. Q. Larson, Bradley D. Hauer, and John D. Teufel
    \vspace{5ex}
\end{center}
\twocolumngrid

\section{3D Cavity}
\label{supp:cavity}
The 3D cavity that houses the readout mode is machined from 6061 Al. The cavity is a rectangular recess with dimensions $l_1=\SI{3.18}{\milli \meter}$, $l_2 =\SI{3.56}{\milli \meter}$, $l_3=\SI{4.06}{\milli \meter}$ and internal edges filleted with a radius of \SI{0.8}{\milli \meter}. The direction of the electric field of the lowest order mode points along the direction of the shortest dimension. We orient the qubit along this direction to generate transmon-cavity coupling. The cavity is constructed from two halves, divided so that currents of the fundamental mode do not cross the seam. The two halves also function to clamp the qubit chip in place. We use a small amount of indium between the chip and one half of the assembly to ensure proper contact. Both halves of cavity are thermalized to the base plate of the cryostat.

It is imperative to simulate the entire 3D structure, including the cavity, sapphire substrate, and transmon features, to accurately model the device. To highlight the importance of including all relevant features in the simulation, we estimate the frequency of fundamental mode of an empty cavity with the dimension used in this work to be $\omega_{r}^{\mathrm{empty}} = 2 \pi \times \frac{c}{2} \sqrt{\left( 1/l_2 \right)^2 + \left( 1/l_3 \right)^2} = 2\pi \times \SI{56}{\giga \hertz}$. In simulation, we find the empty cavity (with filleted internal corners) frequency to be $\omega_{r}^{\mathrm{empty}} = 2\pi \times \SI{58}{\giga \hertz}$. This is more than $\SI{20}{\giga \hertz}$ higher than the simulated mode frequency that includes the relevant components and the measured mode frequency. Unlike typical circuit QED systems, where the 3D cavity mode is between $\qtyrange[range-units=single, range-phrase=\,-\,]{6}{10}{\giga \hertz}$, the sapphire chip and transmon features are no longer small perturbations on the electric field of the mode and must be considered in the design process.

\section{Transmon Fabrication}
\label{supp:fab}
We fabricate the transmon qubit using standard lithography techniques. We start with a C plane sapphire wafer (0001 orientation) from CrysTec. We begin by solvent cleaning the wafer in acetone, isopropyl alcohol, and water. We anneal the wafer in air at \SI{1200}{\degreeCelsius} for 6 hours and let the wafer passively cool to room temperature over the course of 6 hours. We then evaporate \SI{100}{\nano \meter} of 6N Al using an Angstrom electron beam deposition tool. We coat the wafer in SPR660 photolithography resist and pattern the large features ($>\SI{3}{\micro \meter}$) of the device using a MLA 150 maskless aligner. The resist is developed in MF-26A (tetramethylammonium hydroxide based). The exposed Al is wet etched in MF-26A. We strip the remaining resist using acetone and perform a solvent clean.

We apply a multi layer coating in preparation for electron beam lithography. The first layer is HMDS-P20 to promote adhesion, the second is PMGI for the undercut layer, the third is PMMA for the imaging layer, and the fourth is Electra 94 to assist in discharging the sapphire wafer during exposure to the electron beam. We use a \SI{100}{\kilo \electronvolt} JOEL electron beam lithography tool to pattern the sub-\unit{\micro \meter} Josephson junction features. We strip the Electra in water. We independently develop the imaging and undercut layer. First we use a solution of isopropyl alcohol and water in 3:1 ratio by volume at \SI{3}{\degreeCelsius} to develop the imaging PMMA resist. Then we use Developer 101A (tetraethylammonium hydroxide based) to develop the undercut PMGI resist. The undercutting allows us to form a suspended PMMA resist bridge that we use to deposit the Josephson junction. We use an oxygen ash to remove resist residues remaining in the patterned channels.

We load the wafer into an Angstrom angled evaporator, the same electron beam deposition tool used to deposit the base Al metal. We perform an in-situ Argon ion mill to remove oxides where we will make contact between the junction and the rest of the circuit. To create the Josephson junction we first evaporate \SI{35}{\nano \meter} of Al at \ang{45} relative to the wafer normal. Then we fill the chamber with \SI{565}{\milli \torr} of oxygen for 12.5 minutes to form the an aluminum oxide tunnel barrier. We then evaporate \SI{75}{\nano \meter} of Al at \ang{0} relative to the wafer normal to complete the Josephson junction. Finally, we oxide any exposed Al surfaces by filling the chamber with \SI{5}{\torr} of oxygen for 10 minutes.

After junction depositions, we cover the wafer in a layer of protective SPR660 resist. We use a DISCO dicing saw to dice the wafer into individual $\SI{1.5}{\milli \meter} \times \SI{7}{\milli \meter}$ chips, each with a single transmon qubit. Each chip is soaked in PG remover at \SI{80}{\degreeCelsius} to remove the remaining resist and liftoff the excess Al. We finish by cleaning the chips with isopropyl alcohol. For the final step, we subject the chips to highly reactive ozone that strips away remaining residues. Each transmon qubit is probed to measure the resistance of the junction, allowing us to target the desired Josephson energy $E_J$.

\section{Device Parameters}
\label{supp:device_params}
The device parameters can be found in Table \ref{table:device_params}. The transmon transition frequencies are initially measured by spectroscopy and then refined using a Ramsey interferometry measurement. The transmon anharmonicity is the difference between the first two transitions $\alpha_q = \omega_2 - 2\omega_1$.

\begin{figure}
    \centering
    \includegraphics[scale=1.0]{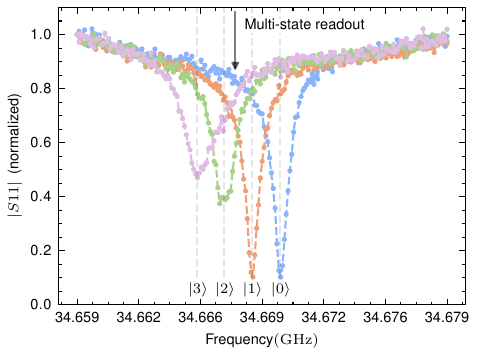}
    \caption{Readout resonator response when the transmon is initialized in $\left\{ \ket{0}, \ket{1}, \ket{2}, \ket{3} \right\}$. The readout resonator is shifted by the cross-Kerr shift $\chi_n$ when the transmon is prepared in state $\ket{n}$. The arrow indicates the frequency used for multi state readout shown in Figure \ref{fig:multistate_readout}.}
    \centering
    \label{fig:cross_kerr}
\end{figure}

The readout cavity is measured in reflection and exhibits a dip the response when the probe is on resonance. The response is fit to extract the readout frequency ($\omega_r$) and linewidth ($\kappa$). Due to higher order nonlinearities of the transmon Hamiltonian, the cross-Kerr interaction rate between the readout mode and a higher level of the transmon, $\ket{n}$, deviate slightly from the simple $n\chi$ relationship described in the main text. In practice, to find the cross-Kerr between the readout resonator and higher levels of the transmon qubit ($\chi_n$), we prepare the state $\ket{n}$ and perform readout spectroscopy. We measure the resonance at $\omega_r + \chi_n$ (Figure \ref{fig:cross_kerr}).

\subsection{Engineering Readout-Transmon Coupling}
For transmons coupled to typical readout modes, in the frequency range of $\qtyrange[range-units=single, range-phrase=\,-\,]{6}{8}{\giga \hertz}$, a coupling rate of $2 \pi \times \qtyrange[range-units=single, range-phrase=\,-\,]{50}{150}{\mega \hertz}$ is needed to achieve the typical cross-Kerr of $\chi \sim -2 \pi \times \SI{1}{\mega \hertz}$. Due to the large ratio of the readout and qubit frequencies in this work, we need to drastically increase the coupling rate to achieve the needed cross-Kerr rates. We measure a cross-Kerr shift of $\chi_1 = -2\pi\times \SI{1.515}{\mega \hertz}$ and estimate the coupling rate between the transmon and readout mode to be $g = \frac{(\omega_r-\omega_1)(\omega_r+\omega_1)}{\omega_r} \sqrt{\chi_1/(8 \alpha_q)} \approx \omega_r \sqrt{\chi_1/(8\alpha_q)} = 2 \pi \times \SI{1.3}{\giga \hertz}$ \cite{gely_nature_2018, kurilovich_high-frequency_2025}. The coupling rate required and achieved in this work is 10 times larger than typical rates found in cQED experiments with $g/\omega_1 = 0.4$. This makes millimeter waves an ideal platform to study light-matter interactions in the ultra strong coupling regime where the ratio of the coupling rate to the transmon frequency can approach unity \cite{frisk_kockum_ultrastrong_2019}.

\subsection{Purcell Effect}
The effective decay rate of the qubit due to its coupling is typically expressed as $\Gamma_p = \kappa \left( \frac{g}{\omega_r - \omega_q} \right)^2$. This approximation neglects the coupling of the readout resonator to the transmission line. When the resonator frequency is above the qubit frequency, the approximation tends to overestimate the effective transmon loss rate. The coupling structure provides additional filtering, modifying the typical decay rate. When the readout mode is capacitively coupled to the transmission line, the effect decay rate is shown in Equation \ref{eqn:purcell} \cite{yen_interferometric_2025}.

\begin{equation}
    \Gamma_p = \kappa \left( \frac{g}{\omega_r-\omega_q} \right)^2 \left( \frac{\omega_q}{\omega_r} \right)^3 \left( \frac{2\omega_q}{\omega_q + \omega_r} \right)^2
    \label{eqn:purcell}
\end{equation}

\begin{figure}
    \centering
    \includegraphics[scale=1.0]{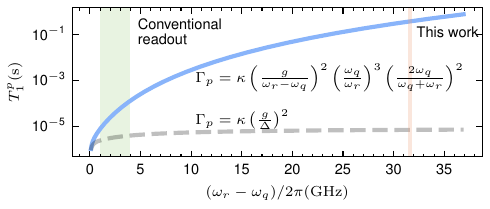}
    \caption{Effective transmon decay time. At large detuning between transmon and qubit, the capacitive coupling between the readout resonator and transmission line provides a significant filtering effect, preventing the transmon from decaying through this loss channel.}
    \centering
    \label{fig:purcell_t1}
\end{figure}

In Figure \ref{fig:purcell_t1} we plot the Purcell limited decay time, $T_1^P$, for a varying readout frequency $\omega_r$ while fixing the qubit frequency $\omega_q$, readout decay rate $\kappa$, and dispersive shift $\chi$ to the values used in this work, see Table \ref{table:device_params}. Since $\chi$ depends on the coupling, readout frequency, and qubit frequency, as shown in Equation \ref{eqn:chi}, at every readout frequency, we compute the $g$ required to obtain the target $\chi$ \cite{kurilovich_high-frequency_2025}. Note the standard expression for the cross Kerr, $\chi = \frac{\alpha g^2}{(\omega_r-\omega_q)(\omega_r-(\omega_q-\alpha))}$, underestimates the shift per photon at large detunings. This allows us to compare the effective decay times across various configurations on an even footing.

\begin{equation}
    \chi = \frac{8 \alpha g^2 \omega_r^2}{(\omega_r^2-\omega_q^2)(\omega_r^2-(\omega_q+\alpha)^2)}
    \label{eqn:chi}
\end{equation}

For conventional detunings of $\SI{1}{\giga \hertz} < \omega_r-\omega_q < \SI{4}{\giga \hertz}$, the decay time is limited to between $\SI{8}{\micro \second} < T_1^P < \SI{130}{\micro \second}$. This is comparable to typical internal decay time of transmons $T_1 \approx \SI{50}{\micro \second}$ and well below the state of art, where $T_1>\SI{500}{\micro \second}$. At these detuning, additional filtering structures, typically an additional resonator, are required to protect the qubits from this decay channel. For large qubit-resonator detunings of $\SI{25}{\giga \hertz} < \omega_r-\omega_q < \SI{35}{\giga \hertz}$, the effective decay times exceed $T_1^P>\SI{150}{\milli \second}$ and can approach $T_1^P \approx \SI{1}{\second}$, much greater than the state of the art transmon decay times. In our device, the strongest form of qubit decay due to external coupling is not the Purcell effect, but rather the direct coupling between the probes and the qubit. Millimeter waves allow us to strongly couple transmons to readout resonators while maintaining high coherence, without the aid of additional filtering structures.

\subsection{Design Parameters}
We note that the design and simulation process of the circuit QED system implemented in this work is no different than that used for conventional frequency readout. We simulate and extract the impedance of the linear modes corresponding to the transmon and readout using Ansys HFSS, construct the Hamiltonian of the coupled system, and diagonalize using \texttt{qutip} to extract relevant parameters. From this process, we simulate the transmon frequency $(2\pi \times \SI{3.087}{\giga \hertz})$, readout frequency $(2\pi \times \SI{33.8}{\giga \hertz})$, cross-Kerr rate $(2\pi \times \SI{1.515}{\mega \hertz})$, and transmon anharmonicity $(-2\pi \times \SI{140}{\mega \hertz})$ \cite{nigg_black-box_2012, lambert_qutip_2026}.

\begin{table}

		\begin{tabular}{c c c}
		    \hline
		    \hline
		     Device Parameter & Symbol & Value \\     
		    \hline
      $\ket{0}-\ket{1}$ frequency & $\omega_1$ & $2\pi\times \SI{3.083}{\giga \hertz}$ \\
      $\ket{0}-\ket{2}$ frequency & $\omega_2$ & $2\pi\times \SI{6.025}{\giga \hertz}$ \\
      $\ket{0}-\ket{3}$ frequency & $\omega_3$ & $2\pi\times \SI{8.813}{\giga \hertz}$ \\
      Transmon anharmonicity & $\alpha_q$ & $-2\pi\times \SI{141.0}{\mega \hertz}$ \\ 
      Transmon decay time & $T_1$ & $\SI{110(5)}{\micro \second}$ \\
      Transmon Echo time & $T_2^E$ & $\SI{188(12)}{\micro \second}$  \\ 
      Transmon residual population & $\bar{n}_0^q$ & $1.2\times10^{-2}$  \\
            \hline

    Readout frequency & $\omega_r$ & $2\pi\times 34.670$ GHz  \\ 
    Readout linewidth & $\kappa$ & $2\pi\times \SI{1.997}{\mega \hertz}$  \\ 
    $\ket{1}$ cross-Kerr & $\chi_1$ & $-2\pi\times \SI{1.515}{\mega \hertz}$ \\
    $\ket{2}$ cross-Kerr & $\chi_2$ & $-2\pi\times \SI{3.002}{\mega \hertz}$ \\
    $\ket{3}$ cross-Kerr & $\chi_3$ & $-2\pi\times \SI{4.457}{\mega \hertz}$ \\
    Cavity residual population & $\bar{n}_0^r$ & $1.06 \pm0.75 \times 10^{-3}$ \\
    \hline \hline
		\end{tabular}
		\caption{Measured device parameters.}
		\label{table:device_params}
\end{table}

\section{Device Operation}
\begin{figure*}
    \centering
    \includegraphics[scale=1.0]{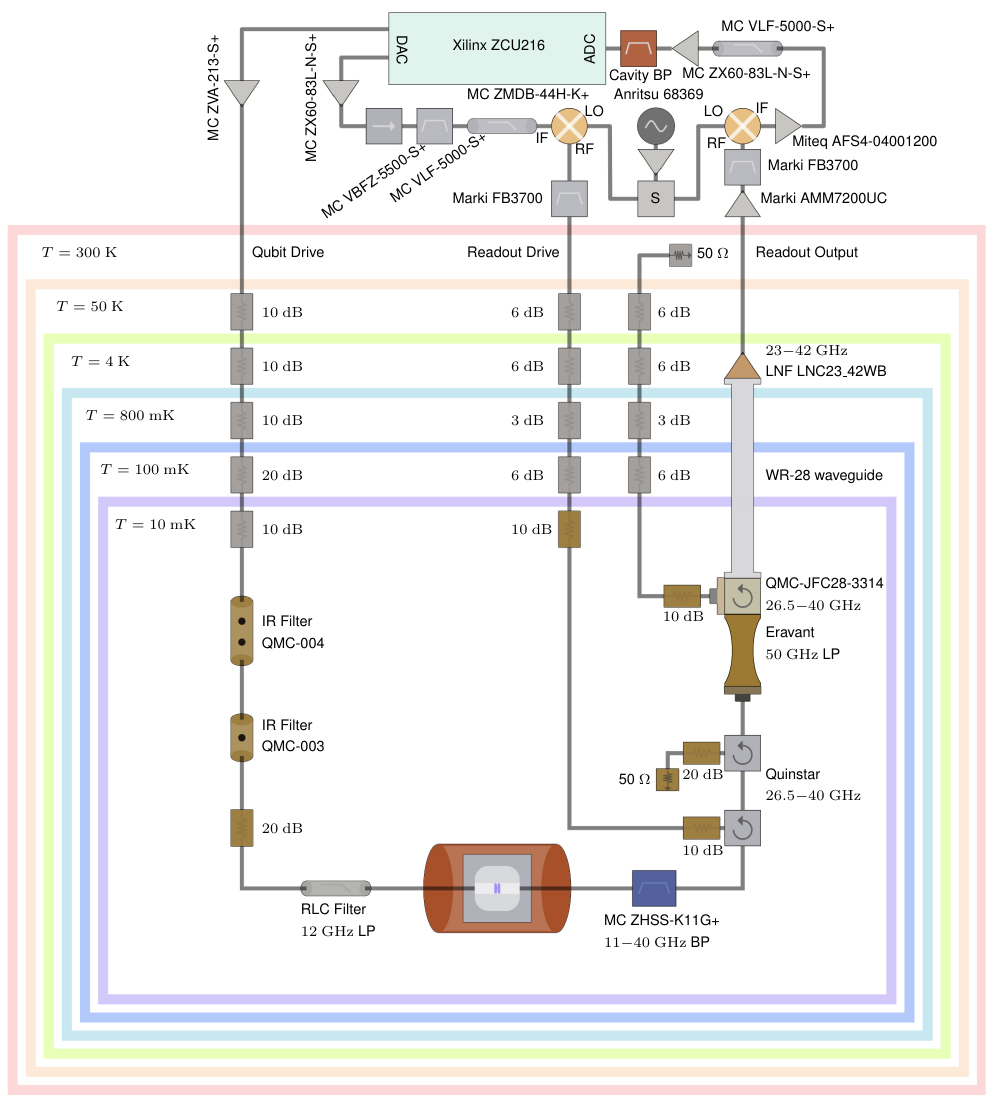}
    \caption{Wiring diagram for millimeter wave readout of a microwave frequency qubit. The qubit drive line is attenuated and filtered to prevent radiation that could populate the qubit or readout cavity or generate quasiparticle excitations in the qubit. The readout drive line has minimal explicit attenuation due to the attenuation inherent in the coaxial lines at millimeter wave frequencies. This line is also filtered to prevent direct qubit excitation and radiation near the band gap of Al from generating quasiparticle excitations. The readout output line has two coaxial circulators followed by a waveguide filter and a waveguide circulator to prevent radiation from traveling down the chain. The signal is carried from the base stage along a waveguide to the $\SI{4}{\kelvin}$ stage where it is amplified by a waveguide HEMT. The qubit drives are directly synthesized by the ZCU216. The readout drive pulses are generated at an intermediate frequency around $\SI{5.67}{\giga \hertz}$ and mixed with a local oscillator at $\SI{29}{\giga \hertz}$. On the input side, the signal is amplified at the IF frequency and then filtered at both the IF and RF frequencies. On the output side, the signal is amplified and filtered at both the RF and IF frequencies. The final filter before the ADC is a mode of a 3D copper cavity that we tune to be in resonance with the down converted signal.}
    \centering
    \label{fig:wiring}
\end{figure*}

The transmon and cavity are thermalized to the $\SI{10}{\milli \kelvin}$ base plate of a cryogen free dilution refrigerator. The device is enclosed in a single layer of OFHC copper shielding. We use radiation shields thermalized at the $\SI{10}{\milli \kelvin}$, $\SI{800}{\milli \kelvin}$, $\SI{4}{\kelvin}$, and $\SI{50}{\kelvin}$ stages and a magnetic shield at $\SI{300}{\kelvin}$. There are two probes that enter the cavity volume along the axis of the fundamental mode electric field direction. The more strongly coupled probe delivers readout and drive pulses as well as carries readout information. The more weakly coupled probe delivers transmon pulses. The coaxial lines that carry millimeter wave signals are connectorized with $\SI{2.92}{\milli \meter}$ connectors. See Figure \ref{fig:wiring} for device wiring inside of the cryostat.

We use a RFSoC ZCU216 with QICK firmware to generate transmon and readout drives as well as record the output signals from the device \cite{stefanazzi_qick_2022}. Our transmon drives are directly generated and delivered to the the cryostat. The readout drives are generated at an intermediate frequency around $\SI{5.67}{\giga \hertz}$ and mixed with a local oscillator at $\SI{29}{\giga \hertz}$ with filtering at both the base band and up converted frequencies. The readout output functions analogously, with the signal mixed down to $\SI{5.67}{\giga \hertz}$ with the same local oscillator at $\SI{29}{\giga \hertz}$. We filter and amplify the output signal at both the signal and down converted frequencies before delivering the signal to the ADC for digitization. See Figure \ref{fig:wiring} for the room temperature electronics setup.

Our device is sufficiently anharmonic to to drive the transmon transitions quickly without significant leakage. We use fast square pulses to transfer the transmon population between adjacent levels $\ket{n} \leftrightarrow \ket{n+1}$: $\tau_{\pi_{01}} = \SI{23}{\nano \second}$, $\tau_{\pi_{12}} = \SI{16}{\nano \second}$, $\tau_{\pi_{23}} = \SI{12}{\nano \second}$.
% \TL{Are these square pulses? Is it worth estimating a worst case leakage given the sinc function?}

\section{Experiment Calibration}
\begin{figure}
    \centering
    \includegraphics[scale=1.0]{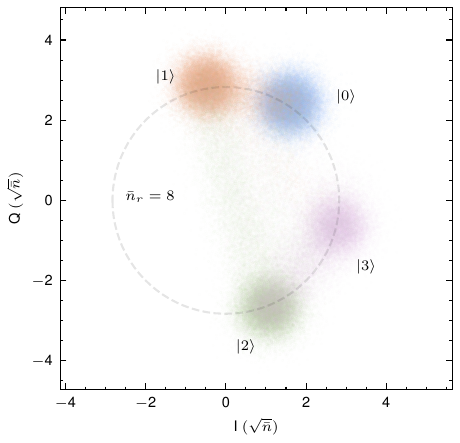}
    \caption{I,Q space distribution of readouts when the transmon is prepared in its first fours states. We choose a readout frequency so that the measured I,Q values for the first four transmon states are distinguishable with $\bar{n}_r=8$ in $\tau_r=\SI{10}{\micro \second}$.}
    \centering
    \label{fig:multistate_readout}
\end{figure}

Measurement of transmon state transition requires the capability to distinguish the various possible states. We use a single readout frequency located between the resonator response with the transmon in 
$\ket{1}$ and $\ket{2}$, $\omega_r^{\ket{2}} < \omega < \omega_r^{\ket{1}}$. At this frequency, we use a drive power corresponding to driving the readout with $\bar{n}_r =8$ photons. The readout is pulsed and the reflected signal is collected for $\tau_r=\SI{10}{\micro \second}$. In Figure \ref{fig:multistate_readout}, we plot the distribution of measured I,Q points with the colors corresponding to the initially prepared state $\ket{n} \in \left \{ \ket{0}, \ket{1}, \ket{2}, \ket{3} \right \}$.

\begin{figure}
    \centering
    \includegraphics[scale=1.0]{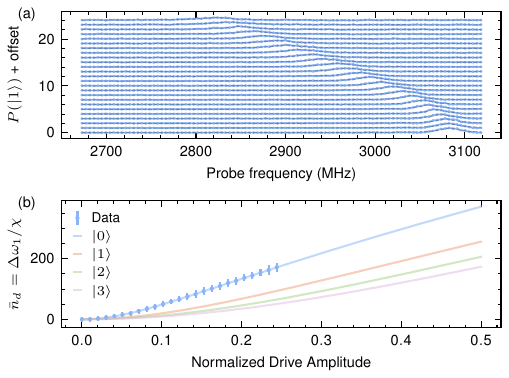}
    \caption{AC-Stark shift calibrates the microwave drive. (a) We measure the transmon $\ket{0}-\ket{1}$ transition frequency as we increase the drive amplitude. The drive photon number is determined from the AC-Stark shift as $\bar{n}_d = \Delta \omega_1/\chi_1$. (b) We fit the relationship between drive photon number $\bar{n}_d$ and normalized drive amplitude $A$ to determine the scaling factor and to extract the readout resonator self-Kerr. We extrapolate the this relationship between drive amplitude and number of drive photons to calibrate the number of drive photons generated at large drive amplitudes and for the qubit prepared in states ${\ket{0}, \ket{1}, \ket{2}, \ket{3}}$. The error bars on the drive photon number $\bar{n}_d$ are the width of the qubit response in (a) scaled by the cross-Kerr shift $\chi_1$. The two contributions to this width are the bandwidth of the probe pulse used ($1/\SI{46}{\nano \second} \sim \SI{22}{\mega \hertz}$) and any measurement induced dephasing of the qubit.}
    \centering
    \label{fig:stark_shift}
\end{figure}

We calibrate the drive strengths used for probing state transitions and performing readout using the AC-Stark effect. The transmon transitions are sensitive to off resonant drives; in Figure \ref{fig:stark_shift}(a) we monitor the first transition whose frequency shifts in response to drives by $\Delta \omega_1$. The AC-Stark shift is proportional to the number of drive photons $\Delta \omega_1 = \bar{n}_d \chi_1$ In Figure \ref{fig:stark_shift}(b) we show the AC-Stark shift as a function of the drive amplitude and fit with the expression in Equation \ref{eqn:n_nonlinear} to calibrate the relationship between drive photon number and drive amplitude.

\begin{equation}
\label{eqn:n_nonlinear}
    \bar{n}_d(A) = C\frac{A^2 \kappa}{(\Delta_d -K \bar{n}_d(A) - n \chi)^2 + (\kappa/2)^2}
\end{equation}
This expression accounts for two effects, qubit state dependence and residual self-Kerr. The resonator response shifts away from our fixed frequency drive when the transmon is initialized in various states, as shown in Figure \ref{fig:cross_kerr}. Due to the coupling between the readout resonator and transmon, the nominally linear mode inherits a small nonlinearity from the qubit. This induces a cavity self-Kerr such that the resonator response shifts as more photons are added. The fit extracts the scaling factor $C$ and the cavity self-Kerr $K$ and we use Equation \ref{eqn:n_nonlinear} to calibrate the drive photon numbers used to measure the transition probabilities presented in this work.

\section{Qubit State Assignment}
\begin{figure}
    \centering
    \includegraphics[scale=1.0]{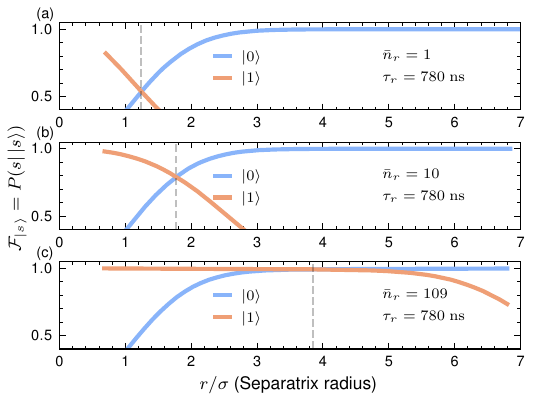}
    \caption{Optimizing the separatrix for state assignment. To distinguish between the two qubit states we use a confining circle centered around the median I,Q value of the $\ket{0}$ distribution. Any point that falls within the circle is labeled at 0 and any point outside is labeled as 1. The radius of the confining circle is optimized to simultaneously maximize the probability of assigning states initialized in $\ket{0}$ to 0 ($P(0 | \ket{0})$) and $\ket{1}$ to 1 ($P(1 | \ket{1})$). We apply this procedure for readout photon number of (a) $\bar{n}_r =1$, (b) $\bar{n}_r =10$, and (c) $\bar{n}_r =109$.}
    \centering
    \label{fig:state_assignment}
\end{figure}

To assess the fidelity of the qubit state measurement we optimize the classification of the measurement result. For a single shot we obtain a I,Q pair ideally corresponding to the qubit state that was initialized. The distributions corresponding to the qubit state typically have a 2D Gaussian profile in I,Q space that we can confine with a circle. To separate the distributions corresponding to the $\ket{0}$ and $\ket{1}$ states, we assign all I,Q values within a circle centered at the $\ket{0}$ distribution to 0 and all values outside the circle are assigned to 1. For a given confining circle radius we calculate the measurement fidelity for each initial state $\mathcal{F}_{\ket{n}} = P(n|\ket{n})$, the probability of assigning the measured I,Q value of the state prepared in $\ket{n}$ to the label $n$. We then optimize the confining radius to jointly maximize $\mathcal{F}_{\ket{0}}$ and $\mathcal{F}_{\ket{1}}$. In Figure \ref{fig:state_assignment} we perform this procedure for three configurations of the single shot readout with $\bar{n}_r \in \left \{ 1, 10, 109 \right \}$. As we increase the size of the confining circle we accurately assign more of the measurements where the qubit was prepared in $\ket{0}$ to 0, increasing $\mathcal{F}_{\ket{0}}$. But, we also mistakenly assign states prepared in $\ket{1}$ to 0, decreasing  $\mathcal{F}_{\ket{1}}$. For each readout photon number, we find the optimal confining circle when $\mathcal{F}_{\ket{0}} = \mathcal{F}_{\ket{1}}$, indicated by the gray dashed line in Figure \ref{fig:state_assignment}(a), (b), (c).

The degree of belief that we have about the state of a qubit, given a readout measurement, is related to the measurement fidelity. After some state manipulation or evolution, we read out the qubit state and assign it to $0$ or $1$. Our goal is to determine which state the qubit is in given the measured data with high credence $P(\ket{n}|n)$. By Bayes rule, this is related to our measurement fidelity:
\begin{align}
    P(\ket{n}|n) &= \frac{P(n|\ket{n}) P(\ket{n})}{P(n)} \nonumber \\
    P(\ket{n}|n) &= \frac{P(n|\ket{n}) P(\ket{n})}{P(n|\ket{n}) P(\ket{n}) + P(n|\ket{\bar{n}}) P(\ket{\bar{n}})} \nonumber
\end{align}
Assuming we have no prior information about the qubit state, $P(\ket{n}) = P(\ket{\bar{n}}) = 1/2$
\begin{align}
    P(\ket{n}|n) &= \frac{P(n|\ket{n})}{P(n|\ket{n}) + P(n|\ket{\bar{n}})} \nonumber \\
    P(\ket{n}|n) &= \frac{\mathcal{F}_{\ket{n}}}{\mathcal{F}_{\ket{n}} + 1-\mathcal{F}_{\ket{\bar{n}}}} \nonumber \\
    P(\ket{n}|n) &= \mathcal{F}_{\ket{n}} \left[\frac{1}{1+ (\mathcal{F}_{\ket{n}} -\mathcal{F}_{\ket{\bar{n}}})}\right]
\end{align}
We see that our degree of belief in the qubit state, given a measurement record, matches the measurement fidelity only when  $\mathcal{F}_{\ket{n}} = \mathcal{F}_{\ket{\bar{n}}}$. When the two fidelities are mismatched, the degree of belief in the qubit state given a measurement can either be greater than (when $\mathcal{F}_{\ket{n}} < \mathcal{F}_{\ket{\bar{n}}}$) or less than (when $\mathcal{F}_{\ket{n}} > \mathcal{F}_{\ket{\bar{n}}}$) expected from the assignment fidelity alone. In general, for a multilevel transmon, we want to minimize the probability of assigning the measurement to the wrong label $\sum_{\ket{\bar{n}}} P(n|\ket{\bar{n}}) \ll P(n|\ket{n})$. In the case of a two level system, this condition manifests as minimizing the difference between the assignment fidelities shown by the dashed lines in Figure \ref{fig:state_assignment}.

\section{Measurement Efficiency Calibration}
\begin{figure}
    \centering
    \includegraphics[scale=1.0]{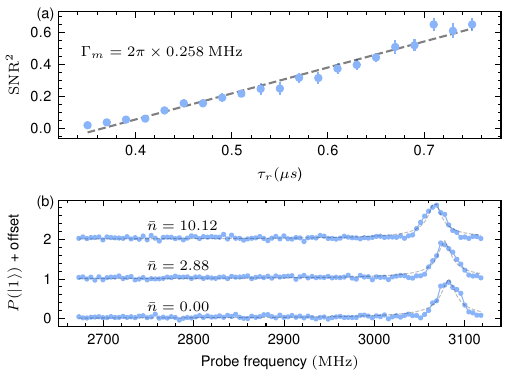}
    \caption{Measurement efficiency calibration. (a) We track the signal to noise ratio of the readout as a function of the readout time to extract the measurement rate $\Gamma_m = \frac{\mathrm{SNR}^2}{4 \tau_r} = 2\pi \times \SI{0.258}{\mega \hertz}$. (b) We use the AC-Stark shift to measure the number of photons in the readout resonator to extract the dephasing rate $\Gamma_{\phi} = \frac{2\bar{n}_r \kappa \chi_1^2}{\kappa^2 + \chi_1^2} = 2\pi \times \SI{14.767}{\mega \hertz}$. The measurement efficiency is given by the ratio of the measurement and dephasing rate $\eta = \Gamma_m/\Gamma_{\phi} = 1.74\%$.}
    \centering
    \label{fig:efficiency}
\end{figure}

We determine the measurement efficiency ($\eta$) as the ratio of the measurement rate ($\Gamma_m$) and the induced dephasing rate ($\Gamma_{\phi}$). We determine the measurement rate (Equation \ref{eqn:meas_rate}) by performing a series of single shot state measurements while varying the readout and integration time.
\begin{equation}
    \Gamma_m = \frac{\mathrm{SNR}^2}{4 \tau_r}
    \label{eqn:meas_rate}
\end{equation}
The SNR is the ratio of the separation of the I,Q distribution when prepared in $\ket{0}$ or $\ket{1}$ to the quadrature sum of the width of the same distributions. In Figure \ref{fig:efficiency}a we fit the relationship between $\mathrm{SNR}^2$ and the readout time $\tau_r$ to a line to extract a measurement rate of $\Gamma_m = 2\pi \times\SI{0.258}{\mega \hertz}$. We determine the induced dephasing rate (Equation \ref{eqn:dephasing_rate}) from the number of photons in the resonator during readout.
\begin{equation}
    \Gamma_{\phi} = \frac{2\bar{n}_r \kappa \chi_1^2}{\kappa^2 + \chi_1^2}
    \label{eqn:dephasing_rate}
\end{equation}
For this particular readout drive strength and frequency we use the AC-Stark shift to measure the readout photon number to be $\bar{n}_r = 10.12$ as shown in the top most curve of Figure \ref{fig:efficiency}b. We calculate an induced dephasing rate of $\Gamma_{\phi} = 2\pi \times \SI{14.767}{\mega \hertz}$. The measurement efficiency is $\eta = \Gamma_m/\Gamma_{\phi} = 1.74 \pm 0.07\%$. The system added noise is computed from the measurement efficiency $\eta = 1/(1+2 \bar{n}_{\mathrm{sys}})$ and we find $\bar{n}_{\mathrm{sys}} = 28 \pm 1$. This is consistent with the added noise from the first state of amplification with our waveguide HEMT $(\bar{n}_{\mathrm{HEMT}} \approx 4)$ and the preceding attenuation from circulators and cabling.

\section{Floquet Simulations}
\label{supp:floquet}
% \begin{itemize}
%     \item floquet simulation of just transmon system. Hamiltonian parameters for Ej Ec determined form fit to experimental data. Highlight a few transitions of note. Blank at 35 GHz
%     \item at high enough power show equivalent to Phys. Rev. Applied 11, 014030 state overlap spread. 
%     \item Inclusion of resonator mode and coupling. Different fit parameters to include Lamb shift and such. Get additional features at half drive frequency. Problematic to simulate right at fr due to ringing up cavity. 
% \end{itemize}

Significant efforts have been made to theoretically and numerically describe the experimentally observed state transitions during dispersive readout \cite{sank_measurement-induced_2016, lescanne_escape_2019, shillito_dynamics_2022, khezri_measurement-induced_2023, cohen_reminiscence_2023, dumas_measurement-induced_2024, nesterov_measurement-induced_2024}. A majority of the effects can be numerically modeled by considering a periodically driven transmon Hamiltonian in the Floquet formalism. The validity of this approximate approach has been studied in \cite{dumas_measurement-induced_2024}, here we concisely describe the particular implementation utilized in this work. 

\subsection{Transmon only system: intrinsic processes}
\label{supp:floquet_transmon}

We follow the approach detailed in \cite{dai_characterization_2026} and utilize the open source package \texttt{floquet} to carry out the numerical Floquet simulations of the system \cite{weiss_floquet_2024}. 

The majority of problematic drive induced state transitions can be accounted for by considering the transmon Hamiltonian with a periodic drive:
\begin{equation}
\label{eq:driven_transmon}
    \hat{H}=4E_C(\hat{n}-n_g)^2-E_J\cos(\hat{\varphi})+E_d\hat{n}\cos(\omega_dt)
\end{equation}

The value of the transmon parameters $E_J$ and $E_C$ are determined by fitting to the experimental qubit frequency and anharmonicity values utilizing the package \texttt{scQubits} \cite{groszkowski_scqubits_2021, chitta_computer-aided_2022}. 
The drive term is a periodic drive of the transmon's charge operator $\hat{n}$ at frequency $\omega_d$. It is important to note inclusion of the offset gate charge term $n_g$ due to the level dispersion of the transmon levels being exponentially more sensitive for higher levels \cite{koch_charge-insensitive_2007}.

\subsubsection{General characteristics}
\begin{figure}
    \centering
    \includegraphics[scale=1.0]{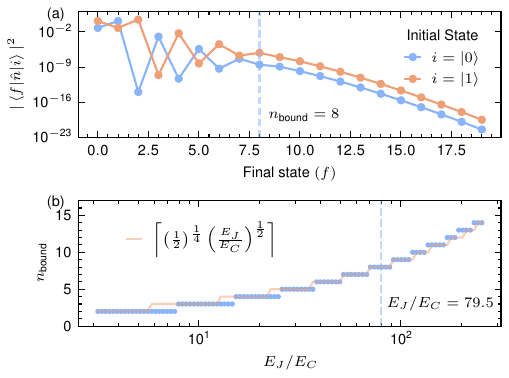}
    \caption{Choosing the ratio of $\omega_r/\omega_1$. (a) The charge matrix elements determine the probability of the transitions between levels. Transitions between levels in the potential well are constrained by parity selection rules, but can be relatively large. Transitions to states outside the well are heavily suppressed. (b) We plot the number of states bound within the well and find that it scales as $\sqrt{E_J/E_C}$. By choosing the readout drive to be $\omega_r/\omega_1>n_{\text{bound}}$ we avoid enabling transitions to states within the potential well that have large charge matrix elements.}
    \centering
    \label{fig:charge_states}
\end{figure}

Under this drive, by Fermi's golden rule, the transition rate is proportional to the matrix element $|\langle f|\hat{n}|i\rangle|^2$ to connect state $|i\rangle$ with $|f\rangle$. In Figure \ref{fig:charge_states}(a) we show a numerical simulation, performed using \texttt{scqubits}, of the charge matrix elements for initial states $\ket{0}$ and $\ket{1}$, with gate charge $n_g=0.25$ \cite{groszkowski_scqubits_2021}. Transitions between states within the transmon potential are strongly limited to states of opposite parity due to the selection rules of the even parity potential. Transition to states ``above'' the cosine potential are not restricted by this selection rule and become exponentially suppressed as the number of excitations increases. 

The goal for choosing a frequency for readout is to avoid a drive frequency that connects low lying levels within the potential because these transitions will have relatively large matrix elements and only a few photons are needed to activate a drive induced transition. Instead, we should choose a readout drive frequency greater than the resonance frequency of either $\left \{ \ket{0}, \ket{1} \right \}$ and the first level outside the cosine potential whose charge matrix element is strongly suppressed as shown in Figure \ref{fig:charge_states}(a). In Figure \ref{fig:charge_states}(b) we calculate the number of states in the cosine well as a function of $E_J/E_C$ with the first transition at fixed frequency at a gate charge of $n_g=0.25$. For the devices used in this work, the first $n_{\mathrm{bound}}=\left \lceil \left( 1/2 \right)^{1/4} \left( E_J/E_C \right)^{1/2} \right \rceil = 8$ levels are bound in the cosine potential. We should therefore choose the readout drive frequency to be $\omega_r/\omega_1 > n_{\mathrm{bound}} = 8$. We choose the readout frequency to be above this threshold $\omega_r / \omega_1 = 11.25$. In addition to being the parameter that sets the transmon's sensitivity to charge noise, $E_J/E_C$ also determines the transmon's susceptibility to drive induced transitions.

% Hence, while there are still states of the transmon that can become on resonance with increased readout drive frequency they are increasingly narrow transitions and do not limit readout performance. Transitions that obey the selection rules can be broad and disruptive to state readout. This is the primary motivation for increasing the qubit-resonator detuning. 

The branch-analysis technique utilized to track and identify the driven transitions with increased drive strength is importantly related to how finely the drive strength is discretized in the numerical simulation.  A finer step size is equivalent to a slower ramping of the drive amplitude, leading to harder to activate transitions being identified in the branch analysis. This is analogous to the adiabaticity of crossing a Floquet connected resonance and has been explored in \cite{wang_probing_2026}. 

The charge dispersion of the transmon energy levels increases exponentially with level number \cite{koch_charge-insensitive_2007} and thus the transition involving higher lying states at a given drive frequency and amplitude can be strongly dependent on the offset charge $n_g$ while transitions involving lower lying states are persistent. In the simulation of Figure \ref{fig:floquet_scars} we varied the offset charge in 11 values evenly spaced from 0.0 to 0.5. 

\subsubsection{Implementation details and definitions}

In the model Hamiltonian the periodic drive is given with amplitude $E_d$. This quantity can not be directly calibrated experimentally. Instead the drive strength can be referred to the AC-Stark shift of the transmon levels for a given drive. Following \cite{dai_characterization_2026} in the simulations we sweep the drive strength via the dimensionless parameter $\xi$:
\begin{equation}
    \xi=\frac{2n_{\text{ZPF}}\omega_d}{\omega_d^2-\omega_q^2}\frac{E_d}{\hbar}
\end{equation}
where $n_\text{ZPF}$ is the zero-point fluctuations of the charge operator, $\omega_d$ is the drive frequency, $\omega_q$ is the transmon's qubit transition frequency, and $E_d$ is the drive strength. Additionally it can be shown that $\xi$ relates the AC-Stark shift of a drive to the anharmonicity 
\begin{equation}
    \Delta_q^{\text{AC}}=\xi^2\alpha_q/2
\end{equation}
which informs an experimentally relevant range to sweep the Floquet simulation drive strength to. 

As the drive frequency and strength is swept in the numerical Floquet simulation the effect of the AC-Stark shift is accounted for by using the concept of ideal-displaced state which maps the bare undriven transmon states to the appropriate driven Floquet mode. Multi-photon resonances can then be identified at a given drive frequency and shape by notable deviation from the ideal displaced state and quantified by an increase in the hybridization:
\begin{equation}
    \Theta(j_t)=1-|\langle\tilde j_t (\xi,\omega_d)|\bar j_t(\xi,\omega_d)\rangle|^2
    \label{eq:hybridization}
\end{equation}
This is the parameter that is plotted in Figure \ref{fig:floquet_scars} and identifies the ``scars'' where the driven transmon state strongly hybridizes with other states. To identify which states participate in a drive-induced resonance, the Floquet modes are iteratively tracked with a state label as the drive amplitude is increased. This technique called branch-analysis gives each Floquet mode an average transmon population which is a weighted sum of overlap with the bare (undriven) transmon number states:
\begin{equation}
    N_i(\xi,\omega_d)=\sum_j j\langle j_t|\tilde i_t(\xi,\omega_d)\rangle
    \label{eq:branch_population}
\end{equation}

As shown in Figure \ref{fig:floquet_scars} of the main text there is a multitude of problematic drive frequencies in the range $\omega_d/2\pi=[4,10]$ GHz, and then additional strong feature around 13 GHz involving the $|0\rangle\rightarrow|5\rangle$ and $|1\rangle\rightarrow|6\rangle$ states. Most notably for increased drive frequency up through the chosen readout frequency around $\omega_d = 2\pi \times \SI{35}{\giga \hertz}$ there is effectively no drive induced state transitions for a transmon around $\omega_q= 2\pi\times \SI{3}{\giga \hertz}$. 
 
\subsection{Transmon-cavity system: mediated processes} \label{supp:cavity_mediated_transition}
The previous section concerned analysis of driven state transitions only concerning the intrinsic transmon Hamiltonian. When there are other modes coupled to the transmon, there can be transitions involving photons across the various subsystems. These can be modeled by a simple extension of the techniques described in the previous section. Extending the Hamiltonian to include a single linear bosonic mode coupled to the transmon's dipole moment can be described by:
\begin{multline}
\label{eq:coupled_floquet}
   \hat{H}=4E_C(\hat{n}-n_g)^2-E_J\cos(\hat{\varphi}) + \omega_a\hat{a}^\dagger\hat{a} \\ -ig\hat{n}(\hat{a}-\hat{a}^\dagger)+E_d\hat{n}\cos(\omega_d t) 
\end{multline}

The drive was chosen to still be associated with the transmon's charge operator, but could equally have been with the cavity mode or some combination of both. This would only lead to a different scaling in relating drive strength to calibrations involving AC-Stark shift. By keeping the drive on the transmon it is consistent with Eq. \ref{eq:driven_transmon} and the drive strength dependence of the results of the previous section. 

\subsubsection{Fundamental readout mode}
It is first instructive to consider the fundamental mode of the readout cavity as the coupled mode \cite{connolly_full_2025, dai_characterization_2026}. The values of $E_J$, $E_C$, $\omega_a$, and $g$ are separately fitted to match the qubit's frequency and anharmonicity and the dressed cavity frequency and dispersive shift. The joint transmon-resonator system is diagonalized using 41 states of the transmon and 5 states of the resonator and we truncate the composite system to the lowest 45 joint system eigenstates. After that, the simulation approach is the same as the transmon only system. 

This simulation captures the same resonant ``scar'' features of the transmon only system and additional features involving both the transmon and readout mode states (denoted $\ket{t,r}$). Previously studied transitions appear around half the readout frequency \cite{chiorescu_coherent_2004, wallraff_sideband_2007, liu_superconducting_2007}. The first involves a coherent exchange of quanta between the transmon and readout mode: $\ket{1,0} \rightarrow \ket{0,1}$ when the drive frequency matches the condition:
\begin{equation}
    2\omega_d=\omega_r-\omega_{01}-\Delta_q^{AC}
\end{equation}
The other visible feature involves generating an entangled state by populating both the transmon and readout mode: $\ket{0,0} \rightarrow \ket{1,1}$ or $\ket{1,0} \rightarrow \ket{2,1}$ with the condition:
\begin{align}
    2\omega_d&=\omega_r+\omega_{01}+\Delta_q^{AC} \\
    2\omega_d&=\omega_r+\omega_{12}+\Delta_q^{AC} 
\end{align}
where the two processes are separated in frequency by the qubit anharmonicity at the same drive power. 

% might be wrong citation here

It is not straightforward to extend this simulation to include the intended drive frequency on resonance with the readout mode. Tracking the ideal displaced states in the $|t,r\rangle$ basis is difficult as the resonant drive will ring up the readout mode. This requires tracing out the readout mode population to recover the transmon state information. This is not of issue, as there are no predicted extra transitions to be concerned of that are mediated by this mode. 

\subsubsection{First spurious mode: experimental verification}
\begin{figure}
    \centering
    \includegraphics[scale=1.0]{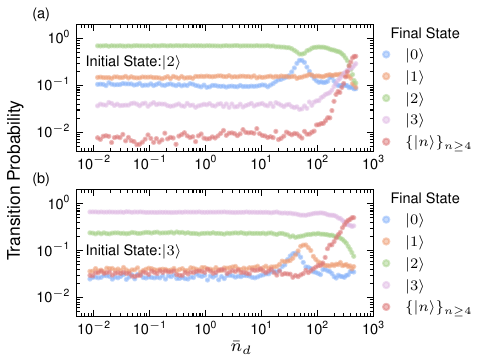}
    \caption{Experimental data probing drive induced transitions of the transmon with $\omega_d/\omega_q=11.25$. We probe the effect of a drive near the readout frequency with varying strength on the transmon states. The transmon is prepared in either (a) $\ket{2}$ or (b) $\ket{3}$ 20,000 times. We track the state of the transmon after applying a variable strength drive in units of drive photon number $\bar{n}_d$ and measure the probability of transitioning from the initial state to various final states. With the transmon initialized in the $\left\{ \ket{2}, \ket{3} \right\}$ subspace, we observe a two-quanta de-excitation process mediated by the first higher order mode of the the 3D cavity. We note that drive photon number calibration is based on the initial prepared transmon state and does not account for decay during the experiment. This leads to the $\ket{2} \rightarrow \ket{0}$ transition seeming to occur at lower drive photon number in (b) as compared to (a) even though the same process is responsible for the transition.}
    \centering
    \label{fig:transition_prob_23}
\end{figure}

We use the fundamental $\SI{34.67}{\giga\hertz}$ mode of the cavity for the transmon readout, but there exist additional linear eigenmodes of the 3D cavity-transmon system. The next lowest eigenmode has a simulated frequency of $\SI{42.7}{\giga \hertz}$ and mediates observable state transitions of the $|2\rangle$ and $|3\rangle$ states. This is seen in Figure \ref{fig:transition_prob_23}(a) where around $\bar{n}_d\approx50$ the initial state of $\ket{2}$ has a transition involving the transmon $\ket{0}$ state and in Figure \ref{fig:transition_prob_23}(b) at slightly lower power there is a $\ket{3} \rightarrow \ket{1}$ transitions.

We can verify with a similar Floquet simulation of Eq. \ref{eq:coupled_floquet} described in the prior subsection where now the linear mode is at $\omega_s/2\pi=\SI{40.0}{\giga\hertz}$, $E_J/h=\SI{10.23}{\giga\hertz}$, $E_C/h=\SI{129}{\mega\hertz}$ and the coupling strength is taken to be $g=2\pi\times\SI{0.5}{\giga\hertz}$. These values are chosen to still give the same qubit transition frequency and anharmonicity. The offset charge is taken to be $n_g=0.25$. The exact choice of $g$ will only affect the linewidth of this transition and is not essential to explain the observed results. We were unable to experimentally verify the exact frequency of this mode due to weak external coupling through the probes and the steep rolloff of the line response from the filtering, circulators, and amplifier.  

The observed transitions involve both the transmon and spurious mode states (denoted $\ket{t,s}$). We measure and simulate the $\ket{2,0} \rightarrow \ket{0,1}$ and $\ket{3,0} \rightarrow \ket{1,1}$ transitions with the condition: 
\begin{align}
    \omega_d=\omega_s-\omega_{02}-\Delta_q^{AC} \\\omega_d=\omega_s-\omega_{13}-\Delta_q^{AC} 
\end{align}

Due to this four-photon mixing process involving one drive photon and one spurious mode excitation at a higher frequency it must involve loss of two excitations of the transmon and thus it only exists for the transmon $|2\rangle$ state and above. These have been studied for the purposes of qubit reset and bosonic mode state preparation and are termed ``f0-g1'' and ``h0-e1'' \cite{pechal_microwave-controlled_2014, magnard_fast_2018}. When the mode in question is spurious, these transitions could be problematic for multilevel qudit operations at these powers.  

% In our experimental system the designed readout mode at $\SI{34.670}{\giga \hertz}$ is the lowest frequency mode above the transmon mode. If there happened to be a lower frequency mode in the environment around $\omega_d-\omega_{02}$ or $\omega_d-\omega_{13}$ this would be a problematic spurious mode for qubit readout.
By operating with millimeter wave photon we are able to suppress intra-transmon resonant transitions. However, this does not eliminate the possibility or inter-transmon or transmon-mode transitions. Consideration for mode mediated processes will be especially important to consider in larger scale implementations of this higher frequency readout strategy. For example, a spurious mode below the readout mode at $\omega_d-\omega_{02}$ or $\omega_d-\omega_{13}$ could mediate $\ket{0} \rightarrow \ket{2}$ or $\ket{1} \rightarrow \ket{3}$ transitions and compromise qubit readout. In 2D architectures, these modes may result from packaging and internal volumes that couple to the qubits. This again highlights incorporating and accounting for all possible structures involved in the system when designing and simulating. For more complex superconducting qubit structures there could be modes inherent to the device that must also be considered \cite{singh_impact_2025}.

\begin{figure}
    \centering
    \includegraphics[scale=0.99]
    {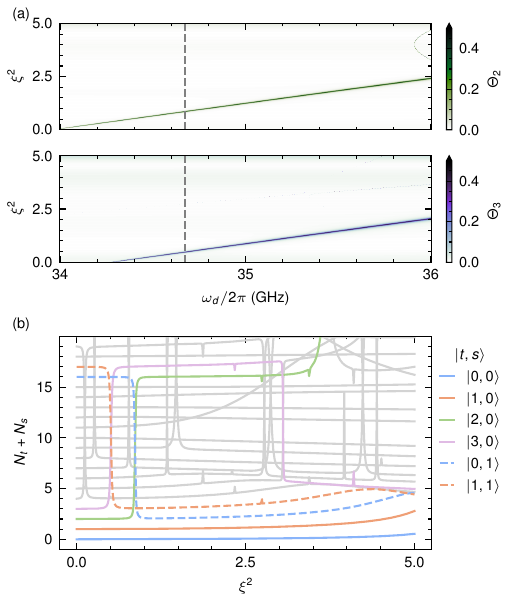}
    \caption{(a) Hybridization of versus drive frequency and drive amplitude for transmon starting in the $\ket{2}$ and $\ket{3}$ states. (b) Branch analysis at $\omega_d=2\pi\times\SI{34.674}{\giga \hertz}$ identifying the two-quanta de-excitation of the transmon with excitation of the spurious cavity mode found in Figure \ref{fig:transition_prob_23}. Average transmon+spurious mode population used to identify branches analogous to Eq. \ref{eq:branch_population}. The simulation verifies the experimental finding that there are no spurious mode mediated transition when starting in the $\ket{0}$ or $\ket{1}$ states as seen in Figure \ref{fig:transition_prob_01}.}
    \label{fig:floquet_spuriousMode}
\end{figure}

% \textbf{Do we need to consider if for some reason the system is starting in $|1_t,1_s\rangle$ or $|0_t,1_s\rangle$?}

% No because this reverse process would be ?
% \begin{equation}
%     \omega_d=-\omega_s+\omega_{02}+\Delta^{AC}
% \end{equation}

\subsection{Transmon escapes confinement at high drive power}
\begin{figure}
    \centering
    \includegraphics[scale=1.0]
    {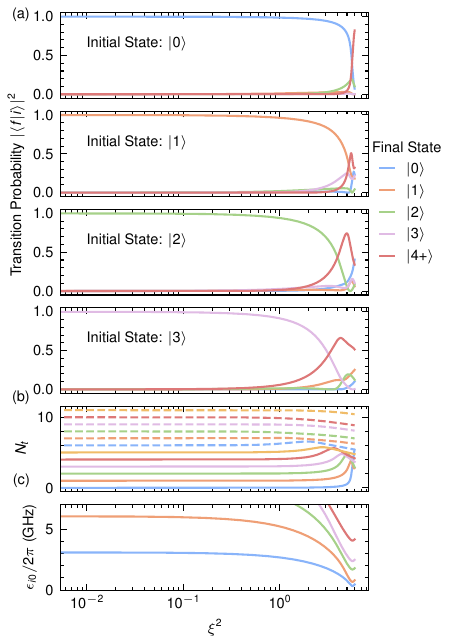}
    % {Figures/highDriveTransition_lin.pdf}
    \caption{(a) Simulated overlap of the Floquet states with the bare transmon eigenstates for different initial state. Analogous to the experimental results of Fig. \ref{fig:transition_prob_01} and \ref{fig:transition_prob_23}. Higher energy initial states deviate at earlier drive strength. (b) Branch analysis under strong drive showing deviation from bare states. (c) Quasienergies of Floquet modes versus drive strength highlighting magnitude of AC-Stark shift. Offset charge $n_g=0.17$. }
    \label{fig:QtoC}
\end{figure}

We are able to drive with extremely strong powers without observing transmon state transitions. Eventually, at high enough powers, we observe the transmon deviate from its prepared state in both experiment (Figure \ref{fig:transition_prob_01}, Figure \ref{fig:transition_prob_23}) and numerical simulations (Figure \ref{fig:QtoC}(a)). This is not a drive induced resonant process, but rather at such high power drive the undriven transmon eigenstates are no longer an appropriate basis of the driven Floquet states (Figure \ref{fig:QtoC}(b)). 

Plotting the overlap of the Floquet states with the bare transmon eigenstates versus drive power we observe similar behavior as described in Reference \cite{lescanne_escape_2019}.%Figure 1(c).
It is interesting to note how much the qubit frequency has been AC-Stark shifted (as computed from the Floquet quasienergies) as shown in Figure \ref{fig:QtoC}(c). 

This coincides with related discussions of readout resonator ``punch-out'' and the ``quantum-to-classical transition'' \cite{reed_high-fidelity_2010, bishop_response_2010, boissonneault_improved_2010, pietikainen_observation_2017}. This provides an upper bound on the drive strength irrespective of resonant transition processes.

We observe in experiment and simulation that the transmon escapes confinement at a lower drive strength when the transmon is prepared in a higher level. This reduction in the upper limit on drive photon number could pose a challenge for high fidelity, QND readout of multilevel qudits or protocols requiring shelving of the transmon state \cite{strauch_quantum_2003, bianchetti_control_2010, elder_high-fidelity_2020}.

